\documentclass{elsart}
\usepackage{natbib}
\usepackage{graphicx}
\usepackage{epsfig}
\begin{document}
%
\def\slasha#1{#1\hskip-0.65em /}  
\def\slashb#1{#1\hskip-1.3em /}   
\def\slashc#1{#1\hskip-.4em /}
%
\def \pb        {{\rm \, pb}}
\def \fb        {{\rm \, fb}}
\def \ipb       {{\rm \, pb^{-1}}}
\def \eV        {{\rm \,  eV}}
\def \keV       {{\rm \, keV}}
\def \MeV       {{\rm \, MeV}}
\def \GeV       {{\rm \, GeV}}
\def \TeV       {{\rm \, TeV}}
\def \MHz       {{\rm \, MHz}}
\def \mrad      {{\rm \, mrad}}
\def \TeVc      {\TeV/c}
\def \TeVcc     {\TeV/c^2}
\def \GeVc      {\GeV/c}
\def \GeVcc     {\GeV/c^2}
\def \MeVc      {\MeV/c}
\def \MeVcc     {\MeV/c^2}
%
\def\ga{\mathrel{\raise.3ex\hbox{$>$\kern-.75em\lower1ex\hbox{$\sim$}}}}
\def\la{\mathrel{\raise.3ex\hbox{$<$\kern-.75em\lower1ex\hbox{$\sim$}}}}
\newcommand{\ckm}{$\checkmark$}
%
\newcommand {\slashed}[1] { \mbox{\rlap{\hbox{/}} #1 }}
\newcommand {\onehalf}    {\raisebox{0.1ex}{${\frac{1}{2}}$}}
\newcommand {\fivethirds} {\raisebox{0.1ex}{${\frac{5}{3}}$}}
\newcommand {\OR}         {{\tt OR}\,}
\newcommand {\BR}         {{\rm BR}\,}
\newcommand {\rts}        {\sqrt{s}}
\newcommand {\lumi}       {\mathcal{L}}
\newcommand {\Lumi}       {\int\lumi\mathrm{d}t}
\newcommand {\gradi}    {^\circ}
\newcommand {\de}         {\partial}
\newcommand {\um}         {\mu \rm m}
\newcommand {\nm}         {\rm   nm}
\newcommand {\us}         {  \mu \rm s}
\newcommand {\cm}         {\rm   cm}
\newcommand {\mm}         {\rm   mm}
\newcommand {\m}          {\rm   m}
\newcommand {\g}          {\rm   g}
\newcommand {\km}         {\rm   km}
\newcommand {\V}          {\rm   V}
\newcommand {\T}          {\rm   T}
\newcommand {\kV}         {\rm   kV}
\newcommand {\nA}         {\rm   nA}
\newcommand {\kVm}        {\rm   kV/m} 
\newcommand {\MVm}        {\rm   MV/m} 
\newcommand {\ns}         {\rm   ns} 
%
\newcommand {\gws}        {\mathrm{SU(2)_L \otimes U(1)_Y}}
\newcommand {\sul}        {\mathrm{SU(2)_L}}
\newcommand {\suc}        {\mathrm{SU(3)_C}}
\newcommand {\ul}         {\mathrm{U(1)_Y}}
\newcommand {\uem}        {\mathrm{U(1)_{em}}}
\newcommand {\sigmabar}   {\overline{\sigma}}
\newcommand {\gmunu}      {g^{\mu \nu}}
\newcommand {\munu}       {{\mu \nu}}
\newcommand {\obra}       {\langle 0 |}
\newcommand {\oket}       {| 0 \rangle}
\newcommand {\bra}        {\langle}
\newcommand {\ket}        {\rangle}
%
\newcommand {\LL}         {L^{\alpha}_{\mathrm L}}
\newcommand {\LLd}        {L^{\dagger \alpha}_{\mathrm L}}
\newcommand {\lL}         {\ell^{\alpha}_{\mathrm L}}
\newcommand {\lLd}        {\ell^{\dagger \alpha}_{\mathrm L}}
\newcommand {\ld}         {\ell^{\dagger \alpha}}
\newcommand {\lb}         {\overline{\ell}^{\alpha}}
\newcommand {\lR}         {\ell^{\alpha}_{\mathrm R}}
\newcommand {\lRd}        {\ell^{\dagger \alpha}_{\mathrm R}}
\newcommand {\nuL}        {\nu^{\alpha}_{\mathrm L}}
\newcommand {\nuLb}       {\overline{\nu}^{\alpha}_{\mathrm L}}
\newcommand {\nub}        {\overline{\nu}^{\alpha}}
\newcommand {\lept}       {\ell^\alpha}
\newcommand {\neut}       {\nu^{\alpha}}
\newcommand {\nuLd}       {\nu^{\dagger \alpha}_{\mathrm L}}
\newcommand {\Phid}       {\Phi^\dagger}
%
\newcommand {\up}         {u^{\alpha}}
\newcommand {\ub}         {\overline{u}^{\alpha}}
\newcommand {\down}       {d^{\alpha}}
\newcommand {\db}         {\overline{d}^{\alpha}}
\newcommand {\QL}         {Q^{\alpha}_{\mathrm L}}
\newcommand {\QLd}        {Q^{\dagger \alpha}_{\mathrm L}}
\newcommand {\UL}         {U^{\alpha}_{\mathrm L}}
\newcommand {\ULd}        {U^{\dagger \alpha}_{\mathrm L}}
\newcommand {\UR}         {U^{\alpha}_{\mathrm R}}
\newcommand {\URd}        {U^{\dagger \alpha}_{\mathrm R}}
\newcommand {\DL}         {D^{\alpha}_{\mathrm L}}
\newcommand {\DLd}        {D^{\dagger \alpha}_{\mathrm L}}
\newcommand {\DR}         {D^{\alpha}_{\mathrm R}}
\newcommand {\DRd}        {D^{\dagger \alpha}_{\mathrm R}}
\newcommand {\bfell}      {\ell\kern-0.4em
                           \ell\kern-0.4em
                           \ell\kern-0.4em
                           \ell }
\newcommand {\obfell}     {\overline{\ell}\kern-0.4em
                           \overline{\ell}\kern-0.4em
                           \overline{\ell}\kern-0.4em
                           \overline{\ell}}
\newcommand {\bfH}      {\; {\cal H}\kern-0.5em \kern-0.4em
                           {\cal H}\kern-0.5em \kern-0.4em
                           {\cal H}\kern0.1em }
\newcommand {\obfH}     {\; \overline{\cal H}\kern-0.5em \kern-0.4em 
                           \overline{\cal H}\kern-0.5em \kern-0.4em 
                           \overline{\cal H}\kern0.1em }
%
\def \b             {{\mathrm b}}
\def \t             {{\mathrm t}}
\def \charm         {{\mathrm c}}
\def \d             {{\mathrm d}}
\def \u             {{\mathrm u}}
\def \e             {{\mathrm e}}
\def \q             {{\mathrm q}}
\def \g             {{\mathrm g}}
\def \p             {{\mathrm p}}
\def \s             {{\mathrm s}}
\def \y             {{\mathrm y}}
\def \n             {{\mathrm n}}
\def \l             {\ell} 
\def \f             {{f}} 
\def \D             {{\mathrm D}}
\def \K             {{\mathrm K}}
\def \Z             {{\mathrm Z}}
\def \W             {{\mathrm W}}
\def \S             {{\mathrm S}}
\def \N             {{\mathrm N}}
\def \L             {{\mathrm L}}
\def \R             {{\mathrm R}}
%
\newcommand {\dm}         {\Delta m}
\newcommand {\dM}         {\Delta M}
\newcommand {\ldm}        {\mbox{``low $\dm$''}}
\newcommand {\hdm}        {\mbox{``high $\dm$''}}
\newcommand {\nnc}        {{\overline{\mathrm N}_{95}}}
\newcommand {\snc}        {{\overline{\sigma}_{95}}}
\newcommand {\susy}       {{supersymmetry}}
\newcommand {\susyc}      {{supersymmetric}}
\newcommand {\aj}         {\mbox{\sf AJ}}
\newcommand {\ajl}        {\mbox{\sf AJL}}
\newcommand {\llh}        {\mbox{\sf LLH}}
%
\newcommand {\rpc}     {{\rm RPC}}
\newcommand {\rpv}     {{\rm RPV}}
\newcommand {\sfe}     {{\tilde{f}}}
\newcommand {\sfL}     {{\tilde{f}_{\mathrm L}}}
\newcommand {\sfR}     {{\tilde{f}_{\mathrm R}}}
\newcommand {\sfone}   {{\tilde{f}_{1}}}
\newcommand {\sftwo}   {{\tilde{f}_{2}}}
\newcommand {\sneu}    {{\tilde{\nu}}}
\newcommand {\wino}    {{\mathrm{\widetilde{W}}}}
\newcommand {\bino}    {{\mathrm{\widetilde{B}}}}
\newcommand {\se}      {{\mathrm{\tilde{e}}}}
\newcommand {\seR}     {{\mathrm{\tilde{e}_{R}}}}
\newcommand {\seL}     {{\mathrm{\tilde{e}_{L}}}}
\newcommand {\st}      {{\mathrm{\tilde{\tau}}}}
\newcommand {\stR}     {{\mathrm{\tilde{\tau}_{R}}}}
\newcommand {\stL}     {{\mathrm{\tilde{\tau}_{L}}}}
\newcommand {\stone}   {{\mathrm{\tilde{\tau}_{1}}}}
\newcommand {\sttwo}   {{\mathrm{\tilde{\tau}_{2}}}}
\newcommand {\sm}      {{\mathrm{\tilde{\mu}}}}
\newcommand {\smR}     {{\mathrm{\tilde{\mu}_{R}}}}
\newcommand {\smL}     {{\mathrm{\tilde{\mu}_{L}}}}
\newcommand {\Sup}     {{\mathrm{\tilde{u}}}}
\newcommand {\suR}     {{\mathrm{\tilde{u}_{R}}}}
\newcommand {\suL}     {{\mathrm{\tilde{u}_{L}}}}
\newcommand {\sdo}     {{\mathrm{\tilde{d}}}}
\newcommand {\sdR}     {{\mathrm{\tilde{d}_{R}}}}
\newcommand {\sdL}     {{\mathrm{\tilde{d}_{L}}}}
\newcommand {\sch}     {{\mathrm{\tilde{c}}}}
\newcommand {\scR}     {{\mathrm{\tilde{c}_{R}}}}
\newcommand {\scL}     {{\mathrm{\tilde{c}_{L}}}}
\newcommand {\sst}     {{\mathrm{\tilde{s}}}}
\newcommand {\ssR}     {{\mathrm{\tilde{s}_{R}}}}
\newcommand {\ssL}     {{\mathrm{\tilde{s}_{L}}}}
\newcommand {\stopR}   {{\tilde{\mathrm{t}}_{R}}}
\newcommand {\stopL}   {{\tilde{\mathrm{t}}_{L}}}
\newcommand {\stopone} {{\tilde{\mathrm{t}}_{1}}}
\newcommand {\stoptwo} {{\mathrm{\tilde{t}_{2}}}}
\newcommand {\sto}     {{\tilde{\mathrm{t}}}}
\newcommand {\SQ}      {{\mathrm{\widetilde{Q}}}}
\newcommand {\STO}     {{\mathrm{\widetilde{T}}}}
\newcommand {\glu}     {{\mathrm{\tilde{g}}}}
\newcommand {\sbotR}   {{\mathrm{\tilde{b}_{R}}}}
\newcommand {\sbotL}   {{\mathrm{\tilde{b}_{L}}}}
\newcommand {\sbotone} {{\mathrm{\tilde{b}_{1}}}}
\newcommand {\sbottwo} {{\mathrm{\tilde{b}_{2}}}}
\newcommand {\sbot}    {{\tilde{\mathrm{b}}}}
\newcommand {\squa}    {{\tilde{\mathrm{q}}}}
\newcommand {\squal}   {{\tilde{\mathrm{q}}_{\rm L}}}
\newcommand {\squar}   {{\tilde{\mathrm{q}}_{\rm R}}}
\newcommand {\sqL}     {{\tilde{\mathrm{q}}_{\rm L}}}
\newcommand {\sqR}     {{\tilde{\mathrm{q}}_{\rm R}}}
\newcommand {\snu}     {{\tilde{\nu}}}
\newcommand {\snue}    {{\tilde{\nu}_{\mathrm e}}}
\newcommand {\snum}    {{\tilde{\nu}_{\mu}}}
\newcommand {\snut}    {{\tilde{\nu}_{\tau}}}
\newcommand {\neu}     {{\chi}}
\newcommand {\chap}    {{\chi^+}}
\newcommand {\cham}    {{\chi^-}}
\newcommand {\chapm}   {{\chi^\pm}}

%
\newcommand {\thstop} {\mathrm{\theta_{\tilde{t}}}}
\newcommand {\thsbot} {\mathrm{\theta_{\tilde{b}}}}
\newcommand {\thsqua} {\mathrm{\theta_{\tilde{q}}}}
\newcommand {\Mcha}{M_{\chi^\pm}}
\newcommand {\Mchi}{M_\chi}
\newcommand {\Msnu}{M_{\tilde{\nu}}}
\newcommand {\tanb}{\tan\beta}
%

%
\newcommand {\rb}    {{\rm R_{\b}}}
\newcommand {\qq}    {{\q \overline{\q}}}
\newcommand {\bb}    {{\b \overline{\b}}}
\newcommand {\ff}    {{\f \bar{\f}}}
\newcommand {\el}    {{\e ^+}}
\newcommand {\po}    {{\e ^-}}
\newcommand {\ee}    {{\e ^+ \e ^-}}
\newcommand {\fbody} {{\sto \to \b \chi {\rm f \bar{f}'}}}
\newcommand {\gaga}  {\gamma\gamma}
\newcommand {\ggqq}  {\gamma\gamma \rightarrow \q\overline{\q}}
\newcommand {\ggtt}  {\gamma\gamma \rightarrow \tau^{+}\tau^{-}}
\newcommand {\qqg}   {\q\overline{\q}\gamma}
\newcommand {\ttg}   {\tau^{+}\tau^{-}\gamma}
\newcommand {\wenu}  {{\rm We\nu_\e}}
\newcommand {\gsZ}   {\gamma^\star\mathrm{Z}}
\newcommand {\ggh}   {\gamma\gamma\rightarrow{\mathrm{hadrons}}}
\newcommand {\ZZg}   {\mathrm ZZ^{*}/\gamma^{*}}
%
\newcommand {\zo}      {{z_0}}
\newcommand {\ip}      {{d_0}}
\newcommand {\thr}     {{{\rm thrust}}}
\newcommand {\athr}    {{\hat{\rm a}_{\rm thrust}}}
\newcommand {\ththr}   {{\theta_{\rm thrust}}}
\newcommand {\acol}    {{\Phi_{\rm acol}}}
\newcommand {\acop}    {{\Phi_{\rm acop}}}
\newcommand {\acopt}   {{\Phi_{\rm acop_T}}}
\newcommand {\thpoint} {\theta_{\rm point}}
\newcommand {\thscat}  {\theta_{\rm scat}}
\newcommand {\etwelve} {E_{12\gradi}}
\newcommand {\ethirty} {E_{30\gradi}}
\newcommand {\eiso}[1] {E^{\, \triangleleft 30\gradi}_{#1}}
\newcommand {\phimiss} {{\phi_{\vec{p}_{\rm miss}}}}
\newcommand {\ewedge}  {E(\phi_{\vec{p}_{\rm miss}}\pm 15\gradi)}
\newcommand {\evis}    {E_{\rm vis}}
\newcommand {\etot}    {E_{\rm vis}}
\newcommand {\emis}    {E_{\rm miss}}
\newcommand {\mvis}    {M_{\rm vis}}
\newcommand {\mtot}    {M_{\rm vis}}
\newcommand {\mmis}    {M_{\rm miss}}
\newcommand {\mhad}    {M^{\rm ex \, \ell_1}_{\rm vis}}
\newcommand {\mhadtwo} {M^{\rm ex \, \ell_1\ell_2}_{\rm vis}}
\newcommand {\ehad}    {E^{\rm NH}_{\rm vis}}
\newcommand {\epho}    {E^{\gamma}_{\rm vis}}
\newcommand {\echa}    {E^{\rm ch}_{\rm vis}}
\newcommand {\nch}     {{N_{\rm ch}}}
\newcommand {\elept}   {E_{\rm lept}}
\newcommand {\elepone} {E_{\ell _1}}
\newcommand {\eleptwo} {E_{\ell _2}}
\newcommand {\pvis}    {{\vec{p}_{\rm vis}}}
\newcommand {\pmis}    {{\vec{p}_{\rm miss}}}
\newcommand {\thmiss}  {{\theta_{\pmis}}}
\newcommand {\pt}      {{p_{\rm t}}}
\newcommand {\ptch}    {{p_{\rm t}^{\rm ch}}}
\newcommand {\pch}    {{p^{\rm ch}}}
\newcommand {\pz}      {{p_z}}
\newcommand {\ptnoNH}  {{p_{\rm t}^{\rm ex \, NH}}}
\newcommand {\puds}    {{P_{\rm uds}}}
%
%
\newcommand{\brchal}{\cal{B}($\PCha \rightarrow \ell\nu\PChi\ $)}
\newcommand{\M}{M_{2}}
\newcommand{\Mp}{M_{2}}
\newcommand{\sigbg}{\sigma_{\mathrm{bg}}}
\newcommand{\ww}   {\mathrm {WW}}
\newcommand{\zz}   {\mathrm Z\gamma^{*}}
\newcommand{\ewnu} {\mathrm{eW}\nu}
\newcommand{\eez}  {\mathrm {eeZ}}
\newcommand{\gagall}{{\gamma\gamma\rightarrow \ell\ell }}
\newcommand{\Pstaup}{{\widetilde{\tau}_{1}}}
\newcommand{\Pstaul}{{\widetilde{\tau}_{L}}}
\newcommand{\Pstaur}{{\widetilde{\tau}_{R}}}
\newcommand{\mzero}{m_{0}}
\newcommand{\msnu}{M_{\tilde{\nu}}}
\newcommand{\mcha}{M_{\chi^{\pm}}}
\newcommand{\mchi}{M_{\chi}}
\newcommand{\mstau}{M_{{\widetilde{\tau}_{1}}}}
\newcommand{\atau}{A_{\tau}}
\newcommand{\chsnu}{\PCha \rightarrow \ell \tilde{\nu}}
\newcommand{\chstau}{\PCha \rightarrow \tilde{\tau}_{1}\nu}
\newcommand{\chlep}{\PCha \rightarrow \ell\nu\chi}
\newcommand{\Tcsq}{\mathrm{TeV}/c^2}
\newcommand{\nobs}{N_{\mathrm{obs}}}
\newcommand{\nlim}{N_{\mathrm{lim}}}
\newcommand{\Brl}{\cal{B}_{\ell}}
\newcommand{\leff} {\mathcal{L}_{\mathrm{eff}}}
\newcommand{\dedx}{{\mathrm{d}}E/{\mathrm{d}}x}
\newcommand{\chtau}{\PCha \rightarrow \tau\nu\chi}
\newcommand{\ssqtw}{\sin^{2}\theta_{\mathrm W}}
\newcommand{\nnz}{{\mathrm \nu\bar{\nu}Z}}
\def \ggll    {\gamma\gamma \rightarrow \ell^{+}{\ell}^{-}}
\def \tautau  {\mathrm \tau^{+}\tau^{-}}
\def \ffg  {f\bar{f}(\gamma)}
\def \lll   {\ell^{+}{\ell}^{-}}
\def \ww   {\mathrm WW}
\def \zz   {\mathrm Z\gamma^{*}}
\def \znn  {\mathrm Z\nu\nu}
\def \zee  {\mathrm Zee}
\def \rts  {\sqrt{s}}
\def \mstop {m_{\tilde{\mathrm{t}}}}
\def \msnu  {m_{\tilde{\nu}}}
\def \elow   {E_{12^{\circ}}}
\def \gev    { \, \mathrm{GeV}/\it{c}^{\mathrm{2}}}
\def \gvm    { \, \mathrm{GeV}/\it{c}}
\def \mx     {M_{\mathrm{eff}}} 
\newcommand{\neutr}{\chi}


\def \Zcc           {\Z \to \charm \bar{\charm} }
\def \Zbb           {\Z \to \b \bar{\b} }
\def \decDS         {\D^{*+} \to \D^0 \pi^+}
\def \decsDS        {\D^{*+} \to \D^0 \pi^+_s}
\def \deckp         {\D^{0} \to \K^- \pi^+}
\def \deckppp       {\D^{0} \to \K^- \pi^+ \pi^+ \pi^-}
\def \deckpp        {\D^{0} \to \K^- \pi^+ \pi^0}
\def \deckpS        {\D^{0} \to \K^- \pi^+ (\pi^0)}
\def \decskp        {\D^{*+} \to \pi^{+}_{s} \K^- \pi^+}
\def \decskppp      {\D^{*+} \to \pi^{+}_{s} \K^- \pi^+ \pi^+ \pi^-}
\def \decskpp       {\D^{*+} \to \pi^{+}_{s} \K^- \pi^+ \pi^0}
\def \decskpS       {\D^{*+} \to \pi^{+}_{s} \K^- \pi^+ (\pi^0)}
\def \epsc          {\varepsilon_{\charm}}
\def \epsb          {\varepsilon_{\b}}
\def \pctod         {P_{\charm \to \D^*}}
\def \pbtod         {P_{\b \to \D^*}}
\def \Gbb           {\Gamma_{\b\bar{\b}}}
\def \Gcc           {\Gamma_{\charm\bar{\charm}}}
\def \Gh            {\Gamma_{\mathrm h}}

\runauthor{Giacomo Sguazzoni}
\begin{frontmatter}
\title{CMS Inner Tracker Detector Modules}
\author[Pisa]{Giacomo Sguazzoni\thanksref{X}}
\thanks[X]{On behalf of the CMS TIB/TID consortium.}
\address[Pisa]{Universit\`a degli Studi e I.N.F.N., Pisa}
\begin{abstract}
The production of silicon detector modules that will instrument the
CMS Inner Tracker has nowadays reached $1300$ units out of the
approximately $3700$ needed in total, with an overall yield close to 
$96\%$. A description of the module 
design, the assembly procedures and the qualification tests is
given. The results of the quality assurance are presented and the
experience gained is discussed.
\end{abstract}
\end{frontmatter}

\section{Introduction}

\begin{figure*}[t]
\centerline{\epsfxsize=0.64\textwidth\epsfbox{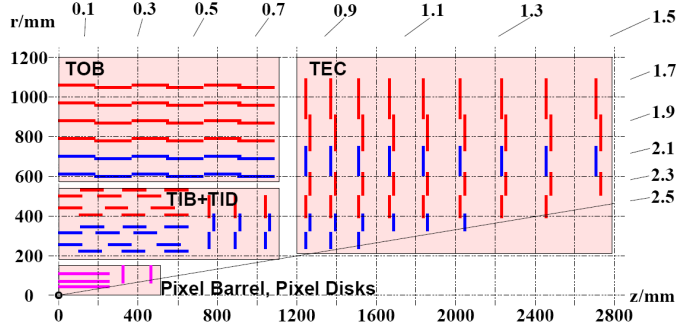}}
\caption{$rz$ cut through one quarter of the SST. The black dot
shows the collision point, the numbers on the top and on the right show the
pseudorapidity coverage of the tracker. Blue/darker lines represents
double-sided layers.\label{fig:layout}} 
\end{figure*}

The CMS Tracker is a cylindrical device of $5.4\m$ in length and
$2.4\m$ in diameter immersed in a 
4 Tesla solenoidal magnetic field. The innermost region ($r<15\cm$) is
occupied by a 
pixel detector, whereas the remaining volume is instrumented by 
using silicon microstrip modules organized in 10 cylindrical
layers and 12 disks as sketched in Fig.~\ref{fig:layout},
corresponding to more than $200\m^2$ of active surface. A detailed
description of the CMS Silicon Strip Tracker (SST) can
be found elsewhere~\cite{TkTDR}. The detector modules described in
this paper will instrument the region comprised between $\sim20\cm$ and
$\sim55\cm$ in radius and $\pm110\cm$ in $z$, i.e. the innermost four
cylindrical layers, known as Tracker Inner Barrel (TIB), and the two groups of
three disks at $|z|$ between $\sim70\cm$ and $\sim110\cm$,
the Tracker Inner Disks (TID), each organized in three concentric
rings. 

Single-sided and double-sided layers can be distinguished in
Fig.~\ref{fig:layout}. Single-sided layers are made of ``$r\phi$''
modules, whose readout strips lay along the $z$ direction for the
barrel and along the radial direction for the
disks, providing a measurement of the $r\phi$ coordinate. The innermost
TIB/TID layers are instead double-sided, i.e. equipped with ``double-sided''
sandwiches capable of a space point measurement and obtained by
coupling back-to-back a $r\phi$ module and a special ``stereo''
module with the strips tilted by $100\mrad$. The TIB requires 2724
modules, 1536 of which assembled into 768 double-sided sandwiches; the
TID requires 816 modules, 576 of which assembled into 288 double-sided
sandwiches. In total 3540 modules must be produced, plus $\sim5\%$ of
spares.

\begin{figure*}[b]
\begin{center}
    \begin{picture}(0,0)
      \put(0,0){\mbox{(a)}}
    \end{picture} 
    \hskip 0.03\textwidth
    \includegraphics[width=0.52\textwidth]{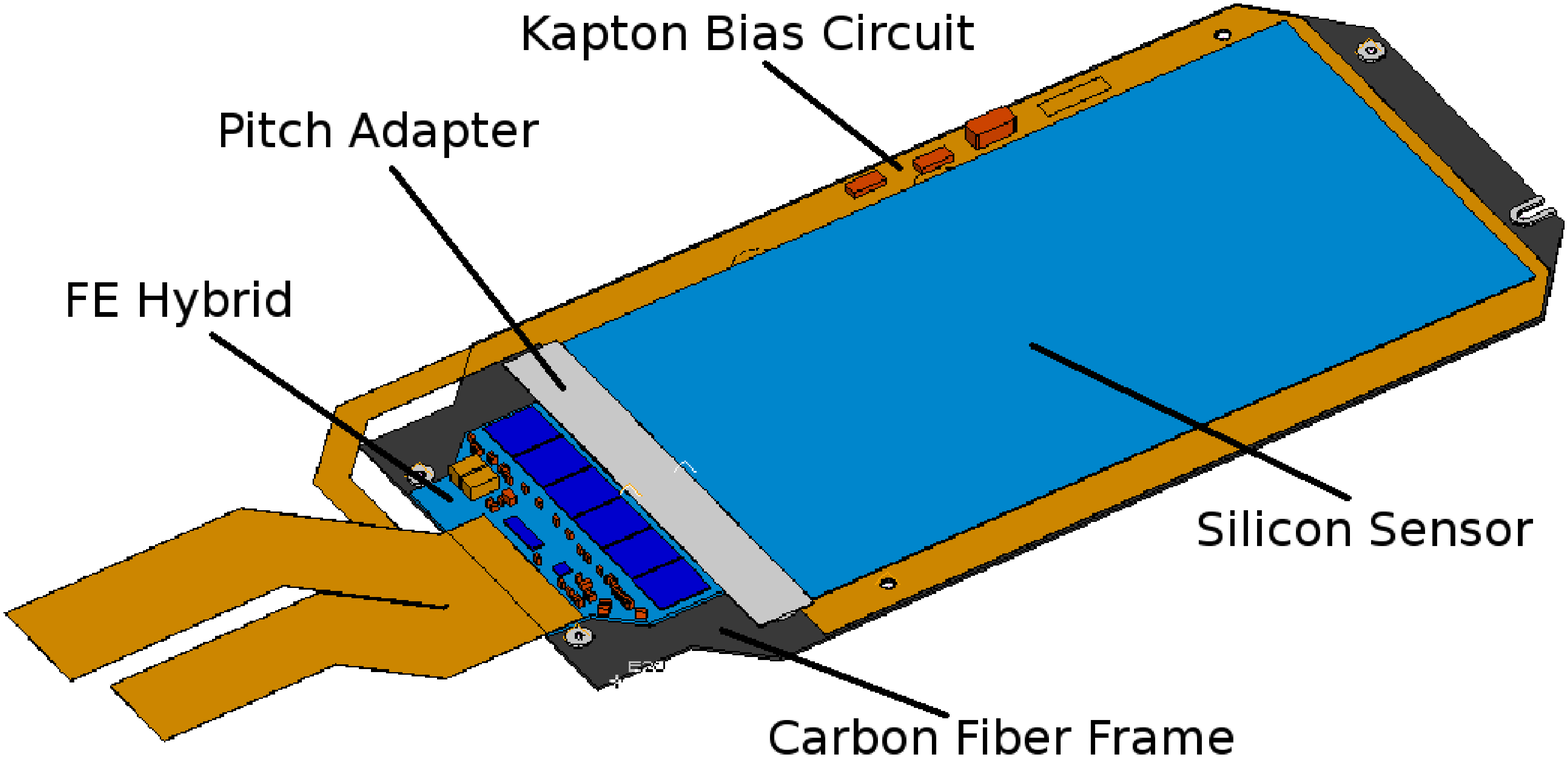}
    \hskip 0.01\textwidth
    \begin{picture}(0,0)
      \put(-14,0){\mbox{(b)}}
    \end{picture}
    \includegraphics[width=0.40\textwidth]{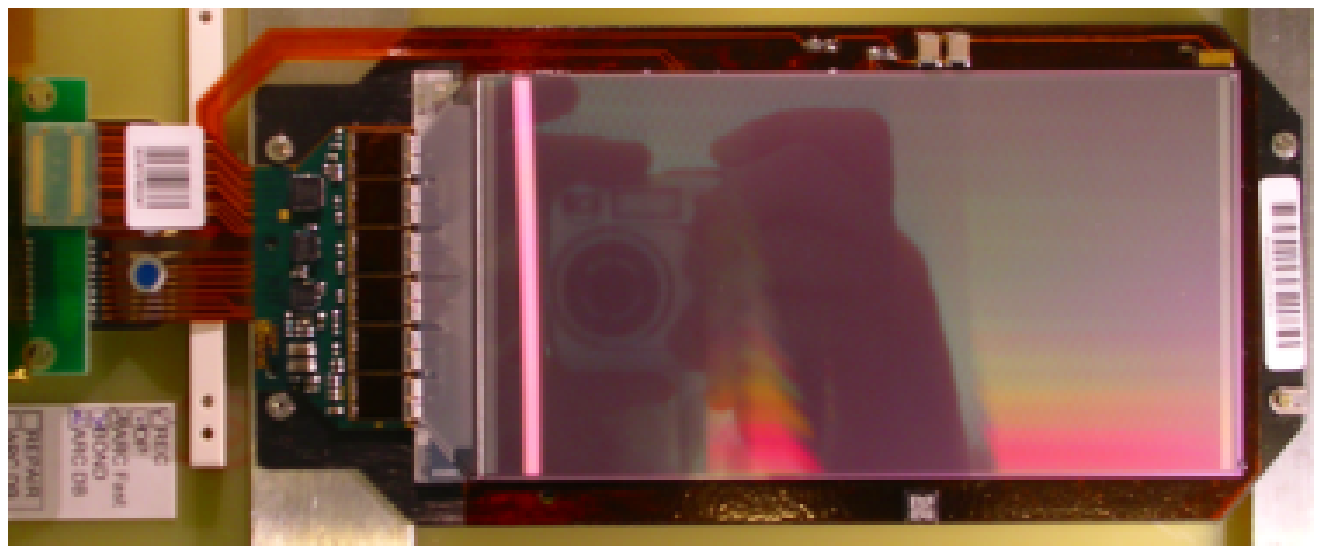}
\caption{(a) Sketch of a TIB barrel module with the various
  components; (b) picture of a TIB barrel
  module.\label{fig:module}} 
\end{center}
\end{figure*}

The TIB/TID modules and their components share the same basic
structure and design with all other SST modules: as shown in
Fig.~\ref{fig:module}(a), the module consists of a carbon fibre frame
that supports the silicon detector and the readout electronics, hosted
on a hybrid. The picture of a module is shown in
Fig.~\ref{fig:module}(b).

All TIB/TID silicon detectors are radiation hard $320\um$ thick
sensors~\cite{sensors} made by Hamamatsu Photonics on 6'' wafer lines. Barrel
sensors are $(120\times62)\mm^2$ with 768 strips at $80\um$ in pitch for
the two innermost layers, and with 512 strips at $120\um$ in pitch for
the outermost layers. Disk detectors have a wedge shaped geometry with
several different dimensions and pitches depending on the ring.

The front-end hybrid is made of a kapton circuit (that also integrates the
connection tail) laminated onto a ceramic rigidifier. It hosts four or six 
APV25, the SST front-end chips~\cite{apv}, the additional ASICs (APVMUX, PLL and
DCU) and the pitch adapter that allows to match the pitch of the APV25
input pads to the pitch of the sensor readout pads.

\section{Module assembly}

Building a high precision silicon tracker device requires that the accuracy
and the reproducibility of the mechanical assembly of its
sub-components lay within few tens of microns, i.e. a figure
comparable with the 
point resolution. This tight requirement is met in our case by means of
semi-automatic assembly gantries that have been designed and
commissioned for this purpose within the SST community~\cite{gantry}. The
``gantry'' is a robot based on a micrometric 3D movement that first 
dispenses the glue on the carbon fibre frame and then places the
components (silicon sensors and front-end  
hybrid) on it. The accuracy is guaranteed by the use of pattern
recognition techniques that exploit the presence of fiducial marks on
each component. At the end of every operation the gantry itself is able to
perform a survey measurement. Typical RMS spread of the relative linear
offset between pieces is below $\sim 8\um$, to be compared with 
acceptance cuts of few tens of microns; the RMS spread of the angular
relative offset is $\sim 4\times10^{-3}\deg$, with an acceptance cut
of $10^{-2}\deg$. Out of more than 1500 modules assembled at the
gantries, less than $1\%$ are rejected because of assembly precision
being out of specifications. 

The connections between the silicon strip sensor readout pads and the
corresponding pitch adapter lines are made by using wire micro-bonding
machines. The bonding of the innermost layer modules with $80\um$
pitch is particularly critical, requiring the bonds to be organized in
two overlapping rows between staggered pads. Bias connections for
the silicon sensor are also made by three groups of micro-bonds, one on the
silicon sensor back and two on the bias ring on the front. Overall,
the TIB/TID project requires more that 2M bonds. Such huge numbers
require the deployment of state-of-the-art micro-bonding
techniques, but this has been certainly achieved since all bonding
centers are performing beyond the specifications: the quality control
procedure, consisting of pull tests on test bonds (and also on
real bonds on a sample basis), have measured an average pull force
exceeding $8\g$, well above the $5\g$ required. More than 0.65M bonds
have already been done up to now, with less than $0.01\%$ of
unrepairable failures.

\section{Quality control tests}

The quality assurance procedure for fully assembled modules consists
of two complementary tests. The ``module full test'' defines the 
module grading by tagging major functional problems and defective 
channels; it follows almost immediately the assembly and bonding
phases for a fast and effective feedback on possible upstream
problems. The ``long term test'' is performed within a climatic chamber
and uses a DAQ system similar to the final one; it is intended to
study possible thermally induced mechanical stresses and module infant
mortality.

The ``module full test'' is performed by using a light, compact,
standalone and user-friendly test system, known as ARC~\cite{arc}. The
test consists of the following measurements: sensor I-V
curve up to 450V; pedestal, noise and pulseshape runs; shorted
neighbouring channel detection by means of cross-talk; open channel
detection by means of LED illumination test, and pinhole detection by
means of light-induced sensor leakage current. A
``open'' is a channel not connected to the 
corresponding sensor readout strip, whereas ``pinhole'' is jargon for
a short-circuited strip coupling capacitor. Shorts and pinholes cause
the lost of the affected channels and, more dangerously, can prevent
an entire readout chip from functioning properly. The standard procedure
requires these channels to be disconnected from the readout electronic
once identified. 

\begin{figure*}[t]
\centerline{
    \includegraphics[width=0.44\textwidth,angle=270]{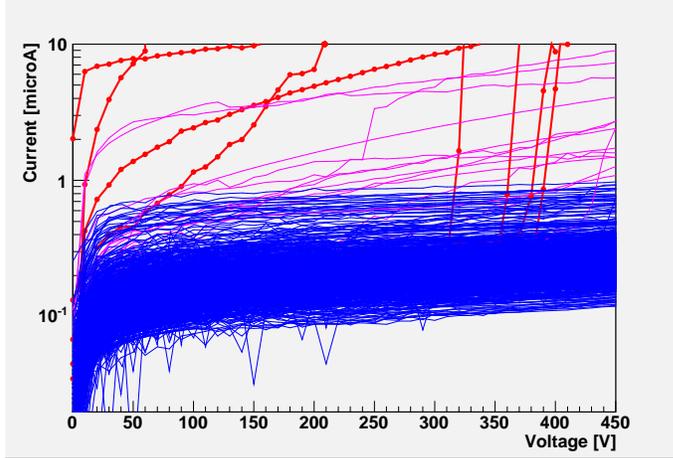}
}
\caption{Sensor currents vs. bias voltage curves for $\sim 1000$ TIB
  modules. Dot-line curves exceed the $10\mu \rm
  A$ acceptance cut at $<450V$ bias voltage.\label{fig:IV}}
\end{figure*}

The current versus voltage (I-V) measurements on the silicon sensors
show a modest degradation with respect to bare sensor
measurements. The I-V curves of $\sim 1000$ modules are shown in
Fig.~\ref{fig:IV}. The current distribution at 450V shows a bulk
around $\sim200\nA$ and only 24 out of more than $1300$ modules ($\sim
1.8\%$) do not comply with the acceptance cut (I(450V)$<10\mu \rm A$).
These failures, concentrated in the early production, have been
traced back to be mainly due to a defective sensor manipulation tool
causing scratches, as later spotted by the detailed optical
inspection that all modules also undergo.

A module is also classified as bad if the number of defective
channels exceed $2\%$ of the total. A channel or strip is considered
bad if flagged as dead, noisy, shorted, open or pinhole. Only $\sim
0.8\%$ of modules have been rejected for this reason up to now: of
over $640$k strips only 1k bad channels have been identified
($\sim 0.15 \%$), approximately half as noisy and half as opens (thus
including also unbonded pinholes and shorts).

The long term test is performed into a climatic chamber, that hosts up to
10 modules, and by which temperature and humidity
are controlled. The test, lasting days, is intended as a real
module burn-in, and consists of the continuous readout of the modules
during multiple thermal cycles between room temperature and $-20^\circ$C
to emulate the real operating conditions within CMS. Over more than a
thousand modules tested, only 11 ($\sim 1\%$) have failed the
long term test. Among them, only one was lost for a suspected
stress-induced failure. The others showed DAQ errors that in all
likelihood are not related to the modules themselves. Most
significantly no new bad channel has been found after the
multiple thermal cycles.

All modules identified as bad, i.e. not complying with the tight
quality requirements at any point of the testing procedure, are now
being collected in a specialized diagnosis and repair center for
further deep investigation and possible recovery.

\section{Conclusions}

After one year of full activity, the TIB/TID community has reached
$\sim40\%$ of the total module production ($1300$ out of $3700$
pieces). The overall yield is stable, around $96\%$. No major
sensor-related failures have been encountered thanks to the excellent
sensor quality and their effective screening. The module
design has been demonstrated to be robust and the
excellent yield proves that the assembly, bonding and testing
procedures are safe and reliable.

\end{document}